\documentclass[conference]{IEEEtran}
\usepackage{cite}
\usepackage{amsmath,amssymb,amsfonts}
\usepackage{algorithmic}
\usepackage{graphicx}
\usepackage{textcomp}
\usepackage{xcolor}

\begin{document}

\title{Decentralized Multi-target Tracking in Urban Environments: Overview and Challenges}

\author{\IEEEauthorblockN{Donald J. Bucci Jr.}
\IEEEauthorblockA{\textit{Lockheed Martin ATL} \\
Cherry Hill, NJ, USA \\
\texttt{donald.j.bucci.jr@lmco.com}}
\and
\IEEEauthorblockN{Pramod K. Varshney} % Add Pranay and Alex as authors...they should help review this
\IEEEauthorblockA{\textit{Syracuse University}\\
Syracuse, NY, USA \\
\texttt{varshney@syr.edu}}
%\and
%\IEEEauthorblockN{Vijay Kumar}
%\IEEEauthorblockA{\textit{University of Pennsylvania}\\
%Philadelphia, PA, USA \\
%\texttt{kumar@cis.upenn.edu}}
}
\maketitle

\begin{abstract}
In multi-target tracking, sensor control involves dynamically configuring sensors to achieve improved tracking performance. 
Many of these techniques focus on sensors with memoryless states (e.g., waveform adaptation, beam scheduling, and sensor selection), lending themselves to computationally efficient control strategies. 
Mobile sensor control for multi-target tracking, however, is significantly more challenging due to the complexity of the platform state dynamics.
This platform complexity necessitates high-fidelity, non-myopic control strategies in order to achieve strong tracking performance while maintaining safe operation.
These sensor control techniques are particularly important in non-cooperative urban surveillance applications including person of interest, vehicle, and unauthorized UAV interdiction.
In this overview paper, we highlight the current state of the art in mobile sensor control for multi-target tracking in urban environments. 
We use this application to motivate the need for closer collaboration between the information fusion, tracking, and control research communities across three challenge areas relevant to the urban surveillance problem.
\end{abstract}

% =========================================================================================
% =========================================================================================
% =========================================================================================

\section{Introduction}
An accurate and scalable multi-target tracking solution is a critical component of many wide-area urban surveillance systems.
For example, human and vehicle detection with closed-circuit television (CCTV) networks leverages multiple bearing-only sensors to uniquely track targets throughout a city \cite{Chen2015,Hou2016,Anuj2017,Yang2009}.
Another important area involves tracking unauthorized unmanned aerial vehicles (UAVs) using heterogeneous and spatially distributed sensors \cite{Poullin2018,Jovanoska2018,Ganti2016}. 
For commercial UAVs that stream video and telemetry data, passive RF detection mechanisms have also been suggested \cite{Watson2017,Watson2016,Scerri2007}.
Across all of these applications, the deployment and positioning of the sensors over time has a major impact on multi-target tracking performance. 
This is especially true when tracking with passive sensors, which requires fusing multiple sensors to unambiguously resolve target position and velocity.
Examples of these passive sensor types include received signal strength indicator (RSSI), time difference of arrival (TDOA), frequency difference of arrival (FDOA), and angle of arrival (AOA).

Sensor deployment and path planning for multi-target tracking falls under the broad research area of \emph{sensor control}.
Sensor control began receiving considerable attention from the information fusion community in the late 1990s \cite{Hero2011, Ng2000}.
Many of the techniques in the area initially focused on dynamic reconfiguration of individual sensors in order to maintain strong target tracking performance (e.g., beam scheduling \cite{Krishnamurthy2001}, waveform selection \cite{Kershaw1994,Sira2009}). 
In the early 2000s, however, the focus shifted to include sensor selection for wireless sensor network (WSN) applications \cite{Ramya2012, Cao2016}.
The size, weight, and power (SWaP) requirements of these systems necessitated sensor control techniques that could balance tracking performance with the energy cost of obtaining and communicating sensor measurements across the network \cite{Guo2004,Zuo2008,Masazade2013,Niu2018}. 
Decentralized multi-target tracking techniques were proposed to maintain communication bandwidth scalability and resilience to sensor failure \cite{Hlinka2013,Hlinka2014,Meyer2018,Msechu2008,Ribeiro2006,Zuo2010}. 
The majority of WSN applications focused on stationary installations, allowing \emph{offline} solutions to the problem of sensor deployment optimization.
A number of WSN deployment optimization solvers were proposed by drawing analogies to the NP-hard art gallery problem from computational geometry \cite{Efrat2005}. 
Meta heuristics formed the basis for many of these solvers, including techniques such as particle swarm optimization and genetic algorithms \cite{Bojkovic2008,Kulkarni2011,Perez2015,Hu2008}.
The sensor control problem for \emph{online} path planning in the context of multi-target tracking, however, is significantly more challenging and less studied.
As opposed to existing offline techniques for path planning in mobile sensor networks \cite{Singh2009}, the multi-target tracking variation of the problem necessitates an online solution due to the lack of \emph{a priori} information on target trajectories.

Very few online mobile sensor control techniques in the current state-of-the-art are capable of addressing the unique challenges associated with non-cooperative target surveillance in urban environments.
This is primarily because the urban environment highly constrains sensor coverage and maneuverability based on terrain elevation and building geometries. 
The same shadowing issues that make target sensing difficult also introduce challenges in maintaining end-to-end network connectivity, thus rendering centralized fusion approaches impractical.
Strong performance and safe operation of a mobile sensor network in this scenario requires an understanding of how the terrain impacts the relevant tracking and sensor control algorithms, all while maintaining decentralized operation.

The goal of this paper is to provide a brief summary of the current multi-target multi-sensor tracking approaches using a mobile sensor network.
We use this summary to highlight the main limitations that prevent immediate application of these architectures to the urban environment.
In the sections that follow, we briefly summarize the model generally assumed for the mobile sensor network control problem. 
Following this, we provide a brief literature review of the current-state-of-the art in multi-target tracking with mobile sensor networks in non-urban environments.
We then conclude by discussing three open challenges related to urban surveillance with commercial-off-the-shelf (COTS) UAVs, or more specifically, quadcopters.

% =========================================================================================
% =========================================================================================
% =========================================================================================

\section{Problem Overview}
\subsection{Integrated Sensing and Control Architecture}
\begin{figure*}
\centerline{\includegraphics[width=0.75\textwidth]{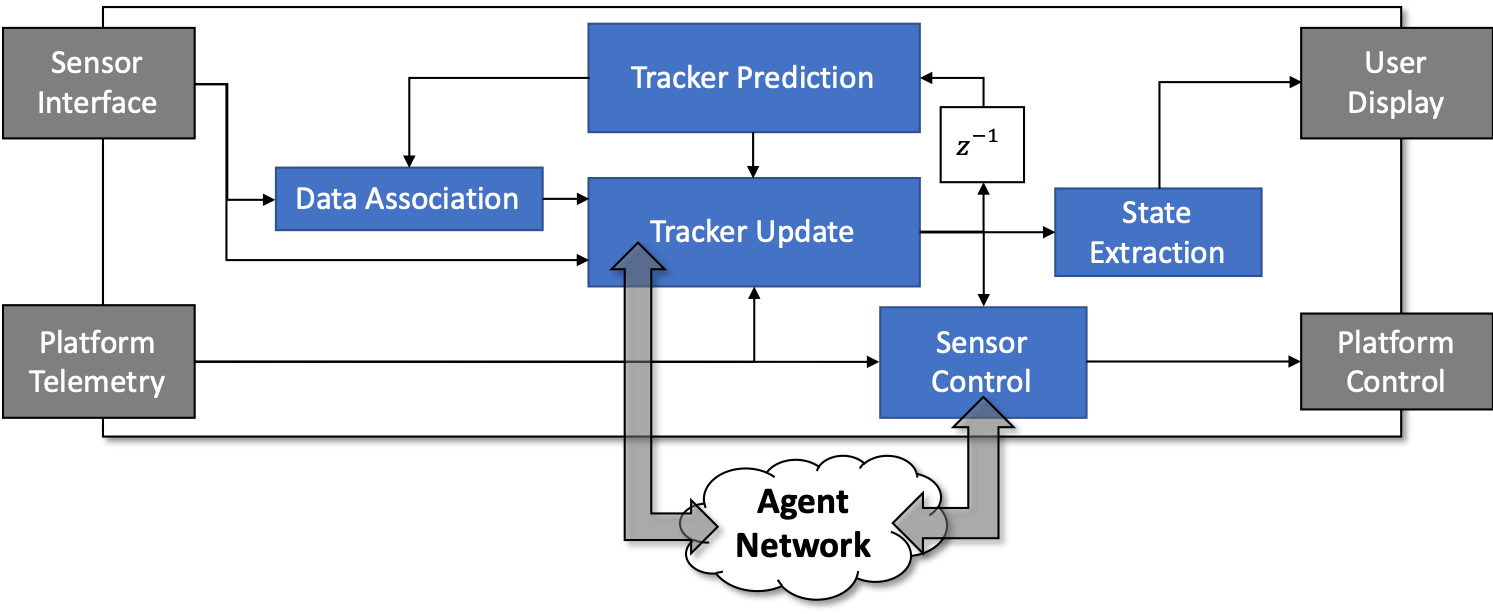}}
\caption{Typical estimation and control architecture used for multi-target tracking. Network communication interfaces shown for decentralized operation. Interfaces shown as gray boxes. Algorithms shown as blue boxes.}
\label{fig:sensingArch}
\end{figure*}

Figure~\ref{fig:sensingArch} shows a typical architecture used for decentralized target tracking and mobile sensor control for a single platform. 
A sensor interface provides derived target measurements, such as time of arrival, Doppler shift, TDOA/FDOA, RSSI, or AOA.
The measurements are usually obtained under measurement origin uncertainty.
That is, it is not known \emph{a priori} which sensor measurements correspond to clutter and which to existing targets.
In addition, measurements obtained from targets may be miss-detected at a given time step.
The posterior distributions for each target from the previous time step are propagated forward in time under known target birth and survival dynamics.
The mechanism for performing this forward prediction is usually a variant of the Chapman-Kolmogorov integral \cite{Sheldon2014,Chen2003}.
A data association process uses these predicted posterior distributions to generate a mapping from the measurements to newborn and persisting targets.
The association map, sensor measurements, and platform telemetry are then used to apply the Bayes update for each target's predicted posterior distribution.
If the tracker update step is decentralized, a consensus process is used to jointly process the sensor log likelihood messages over the network with one-hop neighbors.
The updated posterior distribution per target is used to perform state extraction, which generates state estimates and covariance ellipses.
The sensor control policy finally uses the updated posterior distribution to determine which platform control actions (e.g., heading, acceleration, or waypoint) to use for the next time step.
A separate consensus procedure may also occur to synchronize agent control actions.

% =========================================================================================
% =========================================================================================

\subsection{POMDP Formulation for Mobile Sensor Control}
Control of mobile sensor networks for multi-target tracking typically follows a \emph{partially observable Markov decision process} (POMDP) formulation \cite[Chapter 5.6]{Bertsekas2012}.
In a POMDP, the target states (i.e., position and velocities) evolve according to a Markov process and are observed indirectly through sensor measurements.
Sensor states also evolve according to a Markov process based on the control action applied at the current time step.
Depending on the inertial sensor and kinematic models, the corresponding relationship between platform state dynamics and control may be deterministic or stochastic with directly or partially observable states.
The relationship between the sensor measurements and the target and platform states at the current time step is given by a set of likelihood functions.
The reward function is designed to capture target tracking goals (e.g., minimized cumulative uncertainty in state estimates), obstacle and inter-agent collision avoidance requirements, and constraints on platform control actions.
Given this reward function defined over target states, platform states, and platform actions, the goal is to construct a closed-loop control policy that maximizes the infinite horizon expected cost-to-go (i.e., discounted cumulative reward).
The information available for a control policy at the current time step is the measurement history of all target and platform states and the control inputs used at each platform.
For simplicity, the following discussion will assume that the sensor platform states are completely observable.

To prevent the growth of the control and state space dimensionality as new measurements are obtained, the POMDP model is typically reformulated into an equivalent Markov decision process (MDP).
This is accomplished using a sufficient statistic that subsumes the measurements up until the current time step \cite[Chapter 4.3]{Bertsekas2017}.
The corresponding sufficient statistic is termed the \emph{belief state} of the system.
The belief state for the sensor control application here is the posterior probability of the target states given the observed measurements up until the current time step.
In general, the belief state per target is estimated using application-specific variations of the recursive Bayes filter \cite{Chen2003}.
For the multi-target case, extensions exist for the joint target probability under soft associations \cite{Barshalom1988} or for multi-target probability distribution under the random finite sets (RFS) formalism \cite{Mahler2014}.
The sensor control reward function at each time step is then mapped to the belief states through an information theoretic measure of the quality of the current target state estimate.
The most commonly used measure is the \emph{mutual information} \cite{Cover2006} between the future states per target and the predicted sensor measurements obtained over a finite horizon lookahead window.
Each plausible control action affects the locations of the platforms at future time steps, which in turn affects the posterior belief state for each target.
The core idea is that the mutual information metric quantifies how the the sharpness of the belief state per target changes over a finite horizon lookahead under each control action.

Despite this simplification, the resulting \emph{belief MDP} state space is a subset of the space of multivariate probability distributions.
These belief states are always continuous, even if the partially observable states are discrete.
As a result, very few closed-form optimal policies for belief MDPs exist.
The most well known solution is for the case of a linear Gaussian POMDP with quadratic cost. 
Here, the optimal solution reduces to a Kalman update of the belief state, and the control solution results from solving the discrete-time algebraic Riccati equation \cite{Georgiou2013,Aastrom2012}.
In all other cases, the policies must be determined using approximate online dynamic programming techniques for infinite dimensional state spaces, such as model predictive control (MPC) or stochastic rollout \cite[Chapter 6.4-5]{Bertsekas2017}.
For these techniques, achieving real-time implementation of the control policy involves an application-specific treatment of computational complexity.

% =========================================================================================
% =========================================================================================
% =========================================================================================

\section{Related Work in Non-Urban Environments}
%The main mobile sensor network challenges currently addressed by the community involve the tradeoff between the accuracy of the sensor control problem formulation and the ability to implement non-myopic (i.e., long horizon) control strategies in a computationally efficient manner. 
\subsection{Myopic Control of Mobile Sensors}
Ristic and Vo in \cite{Ristic2010} used the RFS formalism \cite{Mahler2014} to derive a myopic sensor control policy for a single integrator (i.e., velocity controlled) plant. 
The tracking algorithm was a particle filter approximation of the multi-target Bayes filter.
This controller used the Renyi divergence between the predicted and the future expected multi-object posterior after obtaining measurements from range-only mobile sensors.
Ristic et al. provided a similar myopic scheme in \cite{Ristic2011} for range-only tracking, but specialized the multi-object Renyi divergence of \cite{Ristic2010} to the more computationally tractable probability hypothesis density (PHD) filter \cite{Vo2005}.

Gostar et al. in \cite{Gostar2017} leveraged the Cauchy-Schwarz divergence for Poisson point processes \cite{Hoang2015} to derive myopic sensor control policies for a sequential Monte Carlo implementation of the labeled multi-Bernoulli (LMB) filter.
To maximize computational efficiency, the Cauchy-Schwarz divergence between the PHDs of the LMB filter's predict and update steps (per control action) was used as the reward function.
This reward was efficiently realized by evaluating the difference between each predicted and updated particle systems' weights, scaled by each target's probability of existence.
A similar approach was applied by Gostar et al. to the cardinality-balanced multi-target multi-Bernoulli (CB-MeMBer) filter in \cite{Gostar2016}.
To further accelerate computation, both \cite{Gostar2017} and \cite{Gostar2016} applied certainty equivalent (i.e., noiseless) measurement models when predicting future filter states per control action.

Koohifar et al. in \cite{Koohifar2018} provided a single sensor myopic control policy based on the steepest descent direction of the predicted posterior Cramer-Rao lower bound (PCRLB). 
This policy generalized their previous work in \cite{Koohifar2017} by deriving the sensor likelihoods for an RSSI-only measurement.
The RSSI measurement likelihood further modeled packet transmission statistics via a Bernoulli process.
The plant model was assumed velocity controllable, where the heading at each control step was chosen from a fixed quantized set.

Hoffman and Tomlin in \cite{Hoffman2010} leveraged a particle filtering solution and distributed myopic control policy for a constrained double integrator (i.e., acceleration controlled) plant using bearing-only or range-only measurements. 
The plant was designed to model the STARMAC quadcopter \cite{Hoffmann2007} moving at slow speeds. 
The reward function was the mutual information between the predicted future target states and measurements.
To maintain computational tractability, mutual information was evaluated using either local contributions per node, or limited pair-wise contributions between nodes.

Dames and Kumar in \cite{Dames2013} leveraged the PHD filter to construct a distributed sensor control scheme for indoor unmanned ground vehicles (UGV).
In contrast to \cite{Hoffman2010}, the policy used the mutual information between the predicted target states and the binary event of an agent observing an empty measurement set at subsequent time steps. 
Decentralized estimation and control was achieved by transitioning sensors through operating modes, and organizing them into smaller sub-groups where their PHD states were directly synchronized.

Meyer et al. in \cite{Meyer2015} proposed a myopic gradient descent algorithm for a class of decentralized multi-target tracking algorithms based on loopy belief propagation (BP). 
The cost function used was the conditional entropy of the target states at the next time step given the expected sensor measurements at the next time step. 
The relevant BP messages and conditional entropy gradient were approximated using multiple particle systems under perfect knowledge of the number of targets and the target-to-measurement association.
Although the simulations presented in \cite{Meyer2015} were for single integrator dynamics, the corresponding technique is general enough to accommodate non-linear sensor and target dynamic models.

Chung et al. in \cite{Chung2006} proposed a decentralized myopic control strategy based on memoryless, zero cross-covariance, track-to-track fusion. 
The individual sensors provided range-bearing measurements which were fused decentrally using the Kalman equations.
The reward function was the determinant of the fused covariance matrix, which is directly related to the entropy of the fused target state estimate.
The reward gradient was derived, and consisted of a sum of per-sensor reward gradients.
As a result, the control policy was the same for each sensor based on its state, sensor model, and fused target covariance estimates.
A similar controller was derived for the case where imperfect communications contribute to additional errors in the fused estimates.

\subsection{Non-myopic Control of Mobile Sensors}
Beard et al. in \cite{Beard2017} used the generalized labeled multi-Bernoulli (GLMB) filter to apply the Cauchy-Schwarz divergence for Poisson point processes \cite{Hoang2015} as the sensor control reward function. 
The authors also proposed the use of RFS void probabilities to achieve collision avoidance with targets.
An example of controlling of a single range-bearing sensor tracking multiple targets under measurement origin uncertainty was presented.
A finite horizon controller was simulated that assumed a constant velocity plant with instantaneously controllable heading.
Closed form equations of the Cauchy-Schwarz divergence for the case when the GLMB single object posterior densities are modeled as Gaussian mixtures were provided.
The derivations in \cite{Beard2017} for the GLMB can also be applied to the LMB filter as a special case, but exhibit higher complexity than those described by Gostar in \cite{Gostar2016,Gostar2017}.
Implementation details of these control techniques, including pseudo-code, can be found in \cite{Beard2016}

Dames and Kumar in \cite{Dames2014} demonstrated a non-myopic tracking and control solution on real UGVs.
The tracking and control algorithms included a particle filter implementation of the PHD filter and an online estimate of mutual information. 
Similar to \cite{Dames2013}, the non-myopic policy was achieved by evaluating the mutual information against the potential of observing empty measurement sets.
Receding horizon control was achieved through a combination of efficient action-set generation and adaptive sequential information planning \cite{Charrow2014}.

Atanasov et al. in \cite{Atanasov2014} proposed a reduced value iteration (RVI) algorithm demonstrated on a target-linearized range-bearing measurement  model for a single sensor. 
An important distinction was that the relationship between platform state and the observed measurements remained non-linear.
The specific technique did not require linearized sensor platform dynamics, and as such, it was demonstrated in simulation for a single target tracking scenario under differential drive dynamics.
This RVI algorithm was later generalized by Schlotfeldt et al. in \cite{Schlotfeldt2018}  to an anytime planning algorithm.
The resulting technique, denoted Anytime-RVI (ARVI) was decentralized and tested on a set of quadcopters attempting to localize ground-based robots using range and bearing estimates.

Ragi and Chong in \cite{Ragi2013} assumed linear-Gaussian state and measurement dynamics and applied a joint-probabilistic data association (JPDA) tracker \cite{Barshalom1988}.
The sensor control technique used in this paper is known as nominal belief-state optimization (NBO) \cite{Miller2009}.
NBO is a POMDP approach that assumes the associated belief-states (i.e., per target posteriors) are completely characterized by a normal distribution (presumably through a Kalman update).
A certainty-equivalent principle was applied to remove the expectation across belief states.
Both single and multi-step lookahead rollout approaches were provided.
The approach in \cite{Ragi2013} additionally considered forward acceleration thrust and heading dynamics for the platform under wind force disturbances.
Inter-agent collision and obstacle avoidance constraints were considered by including a scaled regularization parameter in the cost-to-go function.

Grocholsky et al. in \cite{Grocholsky2003} assumed a fixed wing aircraft with constant forward velocity and controllable yaw rate to implement a decentralized control rule for bearing-only sensors.
Decentralized data fusion was achieved by leveraging the information form of the Kalman filter \cite{Manyika1994}.
The control law used the expected mutual information gain of the information matrix at the beginning and end of a finite lookahead window.
This law was made computationally feasible by linearizing the measurement and sensor state evolution dynamics and solving the resulting linear-quadratic-Gaussian (LQG) optimal control problem. 

% =========================================================================================
% =========================================================================================
% =========================================================================================

\section{Challenges Specific to Urban Surveillance}
\subsection{Terrain-aware Tracking and Sensor Control}\label{sec:terrain}
The primary sensor control challenge in urban surveillance involves understanding how the terrain and building geometries affect tracking performance. 
In a camera based solution, for example, the observed measurements are AOAs where the probability of detection is dependent on the platform's ability to maintain line-of-sight (LOS) on targets. 
A similar argument can be used for passive RF observations from low power transmitters, where the detectability of multi-path effects is negligible\footnote{Examples of localization in multi-path rich environments have been proposed based off of pattern recognition techniques \cite{Tsalolikhin2011}. These are outside of the scope of this review.}.
If targets maneuver into a non-line-of-sight (NLOS) region to all sensors, the uncertainty on the target's position and velocity increases due to the lack of measurement updates.
Thus, platform maneuvers that keep as many targets within LOS to their corresponding sensors will lead to an increase in the mutual information between predicted future target states and measurements.
Consequently, the number of sensors that have LOS to a given target and their sensing geometry in the LOS region is also important.

There are a number of related studies that provide tracking functionality for targets constrained to road networks.
Ulmke and Koch in \cite{Ulmke2006} describe a particle filtering technique for tracking a single target maneuvering on partially obstructed road networks.
The authors showed that improved tracking performance results when conditioning the measurement detection process on LOS/NLOS information.
Ulmke et al. extended their work in \cite{Ulmke2006} to the RFS formalism in \cite{Ulmke2010} using the Gaussian-mixture cardinalized PHD filter \cite{Vo2007}.
Within these efforts, the detectable regions were constrained by the road network as observed from an overhead sensor.
In the general urban surveillance case, the sensors may not necessarily be overhead.
Furthermore, many practical target types will not be constrained to road networks (e.g., unauthorized UAVs).

The conditioning of the sensor control policy on LOS/NLOS sensing regions as described above necessitates a non-parametric approximation of the belief state per target. 
Across all applications, these approximation techniques are computationally complex and make implementing non-myopic policies at high update rates very challenging. 
A number of point-based value iteration (PBVI) approaches \cite{Kurniawati2008,Smith2005,Kochenderfer2015} have been proposed to solve loosely related target surveillance problems \cite{Egorov2016,Hsu2008,He2010}.
This computational complexity is made worse by the requirement to perform moderate to high fidelity ray-casting under each sensor action to identify the LOS/NLOS regions.
In order to maintain the computational complexity of the PBVI approaches, these shadowing computations necessitate some form of GPU-based acceleration from the computational geometry literature \cite{Tomczak2012}. 

Another important consideration is the incorporation of safety-guaranteed operation with respect to inter-agent and obstacle collision avoidance. 
Some studies such as \cite{Ragi2013} attempt to address the collision avoidance constraints for sensor control by directly penalizing the reward function estimate when targets are too close to other agents or other obstacles. 
A central issue, however, is how the safety constraint penalization term should be weighted when estimating the discounted cost-to-go.
A regularization weight that is too small may not be capable of preventing a collision under the assumed dynamics of the platform. 
Conversely, a penalization that is too large may over constrain the system and unnecessarily degrade the optimality of the policy. 
A better solution would be to select a POMDP solver that is capable of guaranteeing satisfiability of the safety constraints given an accurate map of the environment and agent positions. 
Minimum-norm controllers that modify the planned action from sensor control using safety barrier functions are one option \cite{Ames2016,Freeman1996}, but can potentially over-compensate for safety when the optimization reward is not quantifiable by a control Lyapunov function. 
Computationally tractable implementations of this technique also require a plant model that is affine in the control actions. 
Path planning and graph traversal techniques, such as A* \cite{Hart1968} and RRT \cite{LaValle2001} provide another option, but require a discretization of the platform state space that may not be kinematically feasible. 
Relevant work by the controls community applying such graph search techniques to the trajectory planning problem is given in \cite{Liu2018,Fink2012,Fink2013}.

When digital surface models (DSM) of the terrain and buildings are not available \emph{a priori}, an online estimate is usually computed via a simultaneous localization and mapping (SLAM) technique. 
The use of online estimated map data necessitates sensor control robustness under uncertainty. 
That is, the LOS/NLOS sensor control techniques should be designed to maintain strong performance up to a pre-specified level of error in the estimated map data. 
Similarly, the obstacle avoidance techniques should guarantee collision avoidance up to the same pre-specified level of map error.

% =========================================================================================
% =========================================================================================

\subsection{Control Space Fidelity for Quadcopters}
A major contributing factor to the current interest in urban surveillance with mobile sensor networks is the abundance of commercially available UAVs. 
Quadcopters, for example, provide vertical takeoff and landing functionality in addition to high agility maneuvers.
These platforms and their flexible APIs for flight control tasking \cite{Mavlink,DJIsdk,ROS} make real-time experimentation of mobile sensor network applications very attractive. 
The sensor control methods that exist in the current state-of-the-art, however, make overly simplifying assumptions in the platform kinematics to further reduce computational complexity (e.g., first or second order integrator dynamics).
As a result, the platforms are forced to maneuver at slower velocities so that the actions generated by the sensor control algorithms are representative of the POMDP state dynamics. 
A more critical flaw in this approach is that, under the dynamics of the urban environment, platforms may attempt to delay necessary actions for maintaining collision-free flight until it is too late.

Quadcopter platform dynamics have been studied extensively by the controls and aeronautics communities \cite{Michael2010}.
The quadcopter is a six degree-of-freedom system consisting of position and orientation in 3D Euclidean space. 
However, it provides only four actuation points consisting of the total upward thrust force and the roll, pitch, and yaw moments. 
This makes the quadcopter \emph{underactuated}, implying that its position and orientation can not be accelerated in any arbitrary direction. 
Instead, translational and rotational acceleration are achieved by applying time-varying attitude control.
A naive incorporation of these plant state and action dynamics under a fixed discretization in a POMDP-based sensor control algorithm is not computationally feasible.

Early attempts at quadcopter control applied small-angle approximation techniques to linearize the flight dynamics around the hover state \cite{Hoffmann2008}.  
An important finding was made by Mellinger and Kumar in \cite{Mellinger2011,Mellinger2012}, where the quadcopter was determined to be \emph{differentially flat} in terms of its 3D Euclidean position and yaw angle. 
Differential flatness of a system implies that the original states and inputs can be rewritten as algebraic functions of (potentially fewer) state variables and their derivatives.
These algebraic functions define a diffeomorphism that ensures any trajectories of sufficient smoothness in the flat outputs will be sufficiently smooth in the original state and control space. 

For the quadcopter, the highest degree derivative of the flat position outputs in their expressions for the original control inputs is four (i.e., trajectory snap). 
Similarly, the highest degree derivative of the flat yaw output in the expressions for the original control inputs is two (i.e., yaw acceleration).
Using this insight, Mellinger and Kumar \cite{Mellinger2011,Mellinger2012} provided a series of waypoint-based quadcopter trajectory generation techniques that minimize the control effort under trajectory snap and yaw acceleration (i.e., minimum snap trajectories).
These waypoint generation methods assume a concatenation of piecewise polynomial functions that pass through pre-defined waypoints.
Solving for the trajectory polynomial coefficients is done by solving a computationally tractable quadratic program (QP).
Regulating the original state dynamics of the quadcopter according to this trajectory can then be achieved through the use of a backstepping controller \cite{Michael2010,Lee2010}.

The key takeaway from the above discussion is that the accuracy of the quadcopter control space in a mobile sensor control algorithm can be maintained provided that the actions commanded to the platform generate smooth trajectories up until the fourth derivative of position and second derivative of yaw rate. 
For sensor control with a quadcopter platform, a natural solution is to assume that the plant consists of a fourth order differential equation on the flat outputs, with an input consisting of the trajectory snap at each time step.
This trajectory snap input is termed a \emph{motion primitive} \cite{Liu2017,Liu2018}.
Under polynomial trajectories, these motion primitives induce a resolution-complete discretization in the flat outputs. 
Sikang et al. in \cite{Liu2017} derive this discretization and suggest optimal search techniques for trajectory generation between waypoints using A* \cite{Hart1968}. 
For dynamic environments, Sikang et al. in \cite{Liu2017} proposed a receding horizon control technique based on Lifelong Planning A* (LPA*).
The techniques presented in \cite{Liu2019} were shown to provide collision avoidance guarantees between static and dynamic obstacles, and generate robust paths with respect to random platform disturbances. 

% =========================================================================================
% =========================================================================================

\subsection{Tracking and Control Algorithm Decentralization}
Decentralized operation is a critical requirement of an urban surveillance system.
As discussed in Section~\ref{sec:terrain}, the terrain and building geometries present a strongly RF shadowed propagation environment.
This poses a significant challenge for inter-agent communication, and thus renders centralized tracking and sensor control techniques impractical.
In general, decentralization of multi-target tracking and sensor control algorithms is very challenging
The BP tracking approaches discussed by Meyer et al. in \cite{Meyer2018} provide an intuitive framework for performing average consensus on the relevant belief state parameters with one-hop neighbors.
For particle filtering approaches to the multi-target tracking problem, the BP approaches are decentralized using a consensus-over-weights approach \cite{Farahmand2011}.
Consensus over-weights assumes that the particle systems sampled at each agent are identical, which necessitates the use of synchronized random number generators.
Likelihood consensus \cite{Hlinka2012} is a slightly less restrictive approach that overcomes the need for synchronized random number generators by projecting the sensor likelihood functions onto a common set of basis functions. 
Other alternatives to the consensus-over-weights scheme include fusion via Gaussian mixture approximations \cite{Li2018} and kernel-based methods \cite{Tslil2018}. 

Although these techniques work well when decentralizing target tracking algorithms, it is difficult to apply them to mobile sensor control policies for the urban environment. 
Since the techniques suggested in Section~\ref{sec:terrain} necessitate an online simulation-based approach, it is not immediately clear how a consensus algorithm should be constructed.
One approach to circumvent this challenge is to implement the centralized mobile sensor control policy in a high-fidelity simulation and perform \emph{imitation learning} to generate a decentralized policy. 
Imitation learning is a variation of reinforcement learning, where the goal is to make observations on a set of oracle control decisions and determine a non-parametric representation of the policy \cite{Schaal2003}.
This type of learning has been applied regularly in robotics to learn specific robotic manipulator movements via kinesthetic examples \cite{Kober2010,Pervez2017,Zhang2015}. In these efforts, a convolutional neural network (CNN) is commonly used as approximate architecture for the state-action value function.

A recent study by Gama et al. in \cite{Gama2019} has shown how the convolution and pooling operations used in CNNs can be generalized to support learning with signals supported over graphs. 
The resulting learning architecture, titled a graph neural network (GNN), may be capable of supporting an imitation learning procedure where the one-hop features that may be relevant to consensus are analogous to those signals supported over a communication network graph.
A more thorough investigation of imitation learning of decentralized policies from centralized ones using GNNs is an ongoing and open area of research.

% =========================================================================================
% =========================================================================================
% =========================================================================================

\section{Conclusion}
In this paper, we presented an overview of the mobile sensor control problem for multi-target tracking with a specific emphasis on urban surveillance problems. 
In addition to providing a brief background on the sensor control POMDP formulation, we provided a detailed literature review of the current state-of-the-art and suggested three challenge areas that have yet to receive considerable attention by the community
These three areas were terrain-aware tracking and sensor control, control space fidelity for quadcopters, and joint estimation and control algorithm decentralization.
A number of these challenges are addressed separately in the information fusion, tracking, and control communities.
We suggest a coordinated effort amongst these communities in order to arrive at solutions that are capable of addressing these challenges together in a computationally tractable and bandwidth efficient manner.

% =========================================================================================
% =========================================================================================
% =========================================================================================

\bibliographystyle{IEEEtran}
\bibliography{references.bib}

\end{document}